\documentclass[twocolumn,10pt]{asme2ej}

\usepackage{amsmath}
\usepackage{graphicx}        
\usepackage{subfigure}

\title{Investigation on the performance of a ducted propeller in oblique flow}

\author{Qin Zhang
    \affiliation{Research Fellow\\
	Keppel-NUS Corporate Laboratory\\
	Department of Mechanical Engineering\\
	National University of Singapore\\
	Singapore 119077\\
    Email: mpezhqin@nus.edu.sg
    }	
}

\author{Rajeev K. Jaiman\thanks{mperkj@nus.edu.sg} 
    \affiliation{ Assistant Professor\\
	Keppel-NUS Corporate Laboratory\\
	Department of Mechanical Engineering\\
	National University of Singapore\\
	Singapore 119077\\
	Email: mperkj@nus.edu.sg
    }
}

\author{Peifeng Ma\\
        Research Engineer\\
        Keppel Offshore and Marine Technology Centre\\
        628130,Singapore\\
        Email: PeiFeng.MA@komtech.com.sg\\
       {Jing Liu}
    \affiliation{Research Engineer\\
        Keppel Offshore and Marine Technology Centre\\
        628130,Singapore\\
        Email: jing.liu@komtech.com.sg
    }
}

\begin{document}

\maketitle    

\begin{abstract}
{\it In this study, the ducted propeller has been numerically investigated under oblique flow, which is crucial and challenging for the design and safe operation of thruster driven vessel and dynamic positioning (DP) system. A Reynolds-Averaged Navier-Stokes (RANS) model has been first evaluated in the quasi-steady investigation on a single ducted propeller operating in open water condition, and then a hybrid RANS/LES model is adapted for the transient sliding mesh computations. A representative test geometry considered  here is a marine model thruster which is discretized with structured hexahedral cells, and the gap between the blade tip and nozzle is carefully meshed to capture the flow dynamics. The computational results are assessed by a systematic grid convergence study and compared with the available experimental data. As a part of novel contribution, multiple incidence angles from $15^\circ$ to $60^\circ$ have been analyzed with varying advance coefficients. The main emphasis has been placed on the hydrodynamic loads that act on the propeller blades and nozzle as well as their variation with different configurations. The results reveal that while the nozzle absorbs much effort from the oblique flow, the imbalance between blades at different positions is still noticeable.  
Such unbalance flow dynamics on the blades and the nozzle has a direct implication on the variation of thrust and torque of a marine thruster.
}
\end{abstract}

\begin{nomenclature}
\entry{$u_i$}{The velocity component.}
\entry{$t$}{The time.}
\entry{$\rho$}{The density of fluid.}
\entry{$p$}{The pressure.}
\entry{$p_\infty$} {The static pressure in the free stream.}
\entry{$\nu$}{The kinematic viscosity of fluid.}
\entry{$\theta$}{The angular coordinate.}
\entry{$J$}{The advance coefficient.}
\entry{$J_{BP}$}{The advance coefficient at Bollard Pull condition.}
\entry{$U_\infty$}{The inlet current flow.}
\entry{$n$}{The rotation rate of the propeller, corresponding to the RPM (revolutions per minute).}
\entry{$D$}{The propeller diameter.}
\entry{$\beta$}{The inflow angle.}
\entry{$T_{p}$}{The thrust from propeller component.}
\entry{$T_{n}$}{The thrust from nozzle component.}
\entry{$T_{sp}$}{The thrust from non-ducted propeller.}
\entry{$Q_{p}$}{The ducted propeller torque.}
\entry{$Q_{sp}$}{The non-ducted propeller torque.}
\entry{$\tau$}{The ratio of propeller thrust to total thrust.}
\entry{$K_{TP}$}{The thrust coefficient based on propeller component.}
\entry{$K_{TN}$}{The thrust coefficient based on nozzle component.}
\entry{$K_{QP}$}{The torque coefficient based on propeller component.}
\entry{$\eta$}{The ducted propeller efficiency.}
\entry{$U_{tip}$}{The velocity of the blade tip.}
\entry{$C_p$}{The pressure coefficient.}
\end{nomenclature}

\section{Introduction}

In the recent years, there has been a growing trend of research on ducted propellers (i.e., thrusters) due to their vast industrial applications such as in the DP systems of semi-submersible and marine vessels. To achieve high accuracy in operation and safe stationkeeping, the ducted propeller performances need to be analyzed in a detailed manner under varying conditions, which is crucial and challenging for the success of a DP controlled platform or vessel. For example, during DP operation, thrusters underneath the vessel counteract wave and current forces from different directions. To maintain the predefined position and heading, the ducted propeller is likely to work in the off-design conditions. Unbalanced hydrodynamic forces on blades and nozzle can lead to variation of thrust and torque. The understanding of ducted propeller related hydrodynamic can pave a way for improving the strategy for DP algorithms, which contributes to a safer and optimal marine operations. Furthermore, the reliability and efficiency of the thruster design directly transform into a lower energy cost and a higher performance.

In literature, ducted propeller has been extensively studied  through physical experiments. In \cite{RN1189}, the authors  investigated the in-tubed propellers for aeronautical applications. The authors \cite{RN1186} systematically conducted a series of physical experiments to analyze the characteristics of DP ship thrusters. A principal guideline for the selection of thrusters for dynamic positioning vessels has been proposed by \cite{RN1176}. In \cite{RN972}, the authors summarized the existing experimental data, and proposed several empirical formulas for the momentum loss during the thruster-thruster interaction. Compared to the ducted propeller loading study, the investigation of wake flow of ducted propeller is a relatively unexplored area. In \cite{RN980}, Nienhuis investigated the effects of thruster interaction and measured velocities in the wake flow with LDV (Laser Doppler Velocimetry). PIV (Particle Image Velocimetry) measurement of wake flow has been conducted by \cite{RN1194} at the downstream location up to 1.5 thruster propeller diameter. The most recent physical experiment about thruster wake flow was conducted by MARIN for the configurations such as single thruster in open water conditions, thruster under a plate, thruster under a barge \cite{RN947}, semi-submersible and DP drill ship \cite{RN1019, RN960}. 

Apart from the physical experimental studies, the application of computational fluid dynamics (CFD) to analysis and design of marine propellers has been considered in the last 20 years \cite{RN967}. Multiple models and methods adopted in open propeller investigation have also been extended to ducted propeller-related interaction effects. For instant, vortex lattice lifting-line theory is used to model an asymmetric-ducted propeller with no gap between the duct and the propeller \cite{RN939}. The Multiple Reference Frame (MRF) method as a quasi-steady method is generally used to calculate the ducted propeller performance such as thrust, torque and efficiency \cite{sanchez2000simulation, RN940, RN1103, RN1045, RN1130, RN972, RN954, RN1160, RN1159}. Recently, the scale effect on the open water characteristics of ducted propellers was investigated with the MRF method via a commercial solver \cite{Bhattacharyya_2016,RN1188}. The transient sliding-interface simulation of full blade geometry with a moving mesh was conducted for industrial purpose with commercial software \cite{RN1196, RN1094, RN940}. To investigate the wake flow of ducted propeller and reduce the computational resource requirement, several hybrid models have been developed and explored. One of the examples is the modeling of combined vortex-lattice method (MPUF-3A) with a RANS solver in a commercial code for the unsteady flow analysis to predict the effective wake of thrusters \cite{RN927} and thruster-hull interaction\cite{RN981}. A coupled hybrid approach whereby the actuator disk model for the propeller blade and the RANS model for the ducted propeller wake flow field was explored in \cite{RN990, RN959}. The MRF steady-state simulation was compared with the sliding mesh transient simulation in the case of a single ducted propeller with a rudder by \cite{RN1309}, which showed one percent difference in prediction of forces. In the study of \cite{RN935}, the MRF method in OpenFOAM was evaluated with a marine propeller using simpleFOAM steady state solver, the results agreed very well with the experimental data. It can be noted that as compared to transient sliding-interface technique, the MRF method generally requires much lesser computing resource and provides a reasonable prediction related to the hydrodynamic force. 
The investigation of an open propeller and a ducted propeller has also been extended to the off-design condition. A detail analysis of performances for a marine propeller operating in oblique flow was conducted from $10^\circ$ to $50^\circ$ by \cite{RN1240,RN1241}. In \cite{RN990}, the wake of an azimuthing thruster ($7^\circ$) was investigated physically and numerically. The hydrodynamics of tilted thrusters ($7^\circ$) was investigated by using MRF steady state simulation in \cite{RN1094}. 
There is a lack of systematic investigation dealing with the influence of the off-design condition on a single ducted propeller, which motivates the present 
study. 

This numerical study first investigates the single ducted propeller under a wide range of advance coefficients and inflow angles via quasi-steady MRF-RANS model simulation and transient sliding mesh hybrid LES/RANS model simulation. The results of ducted propeller force are compared with the experimental and the recent numerical data. The propeller blades and nozzle is mainly investigated through the force and pressure, the flow field in front and back of ducted propeller is also revealed. These investigations can be also used in the structural analyses for thruster-driven vessel and DP system to determine the stress and vibrations that occur in the unit under different operating conditions.

\section{Numerical Modeling}
The ducted propeller flow dynamics is governed by the Navier-Stokes equations of incompressible flow posed on a moving frame. For the sake of completeness, we briefly present the RANS model and the underlying numerical details.
\subsection{Governing equations}
The complex flow fields around the blades of the propeller is obtained by solving the three-dimensional Reynolds-averaged Navier-Stokes equations, which can be written in the Cartesian coordinate form as:
\begin{equation}
\frac{\partial {u_i}}{\partial t} + u_j \frac{\partial {u_i}}{\partial {x_j}} = - \frac{1}{\rho} \frac{\partial p}{\partial {x_i}}+ \nu \frac{\partial ^2 {u_i}}{\partial {x_i}^2} + \frac{\partial }{\partial {x_j}}(-\overline{u'_i u'_j})
\label{eq_1}
\end{equation}
\begin{equation}
\frac{\partial u_i}{\partial x_i}=0
\label{eq_2}
\end{equation}
where $u_i$ is the velocity component, $u'_i$ and $u'_j$ are the fluctuating velocities, $t$ is the time, $\rho$ is the density of fluid, $p$ is the pressure, $\nu$ is the kinematic viscosity of fluid, $x_i$ are the coordinates. 
In this study, the turbulence effects are modeled by the one equation eddy viscosity model, namely Spalart-Allmaras (SA) model \cite{RN593} for the steady state simulation. It should be noted that there is a ``trip function" with $f_{t2}$ term in the original S-A model, which is a numerical fix to delay the transitional flow. The implemented S-A model is without the trip-term since there is not much difference with it and most practical applications tend to run the model in fully turbulent mode \cite{RN1122}. Another modification is to prevent the $\tilde{S}$ term from becoming negative for complex flows, here a limiter proposed by \cite{RN1201} is implemented in which $\tilde{S}$ is clipped at $C_s\Omega$ with the default value of $C_s=0.3$.
The SA-based DDES approach proposed by \cite{RN592} is applied for the transient computation with moving mesh.
\subsection{Treatment of rotating fluid-solid interface}
The complex flows past rotating objects can be computed using several approaches. Two flow domains are considered in this study, one with all rotating parts is referenced as ``rotor" domain, the rest is referenced as``stator" domain. In a steady-state simulation, the ``rotor" domain is solved in the rotating frame, while ``stator" domain is solved in the stationary frame, which is well-known the MRF method. 
For the transient simulation, the governing equations for continuity, momentum, turbulence and rigid body motion are solved separately for the rotating and the static domains. Fluid in both domains are coupled across their interfaces using a conservative interpolation method via the local Galerkin projection proposed by \cite{RN975}, namely Arbitrary Mesh Interface (AMI). It is a robust, efficient supermesh construction algorithm, and enables simulation across disconnected and adjacent mesh domains.

\subsection{Numerical discretization}
The governing equations are solved at cell centers in the finite volume domain. For MRF steady-state simulation, temporal derivative contribution for the time scheme is set to zero. The second-order gradient-term discretization and bounded second-order upwind divergence-term discretization are employed. The SIMPLE (Semi-Implicit Method for Pressure Linked Equations) solver given by \cite{RN1315} is used to obtain the steady state solution in the present study.
For the AMI-based transient simulation, the temporal term is implicit with second-order convergence and the gradient term discretization is also second-order accurate. A hybrid convection scheme mentioned by \cite{RN734,RN1117} is adopted for the DDES calculations, which blends between a low dissipative unbounded second order convection scheme in the LES region and a numerically more robust unbounded seconder order upwind scheme in the RANS region. The continuity and momentum equations are solved by the hybrid solver, which combines the time-accurate PISO (Pressure Implicit with Splitting of Operators) algorithm and the SIMPLE algorithm. The system of linear equations resulting from the discretized equations is solved by means of the Gauss-Seidel and the generalized geometric-algebraic multi-grid (GAMG) solvers.

\section{Test Case Description}
\begin{figure}
	\centering
	\subfigure[Front view.]{\label{geo_front}\includegraphics[width=1.6in]{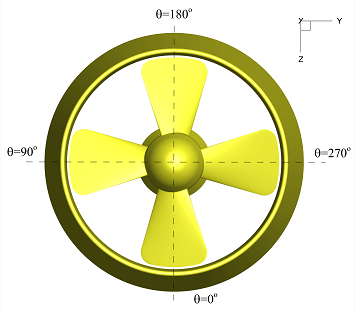}}
	\subfigure[Side view.]{\label{geo_side}\includegraphics[width=1.6in]{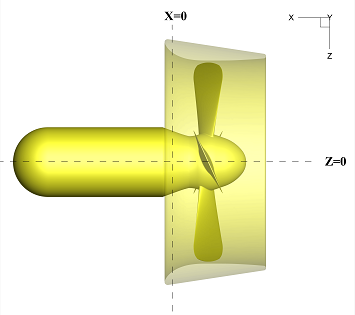}}
	\caption{Geometry of the ducted propeller.The blade position according to angular coordinate is labeled in the left figure.} \label{thruster_geometry}
\end{figure}

As mentioned earlier, the test case chosen for this investigation is a single straight thruster rotating in open water from the recent MARIN Joint Industry Project, \cite{RN1019, RN947}. The experiments were conducted in the MARIN Deepwater Towing Tank, and the thruster properties are listed in Table~\ref{table_2}, the geometry of the thruster model is shown in Fig.~\ref{thruster_geometry} with front view and side view respectively. The propeller reference system defined in \cite{RN1284} is adopted in this study, namely the $X$-axis is positive, forward and coincident with the shaft axis; the $Y$-axis is positive to starboard and the $Z$-axis is positive in the vertically downward direction\cite{RN967}. The angular coordinate $\theta$ originated from $Z$-axis is also labeled in Fig.~\ref{thruster_geometry}. During the physical experiments, the propeller thrust/torque and the nozzle thrust were measured.

\begin{table}[htbp]
	\caption{THRUSTER PARAMETERS \cite{RN947}}
	\begin{center}
		\label{table_2}
		\begin{tabular}{c c c}
			& & \\ 
			\hline
			Parameters         & Unit & Value \\
			\hline
			Propeller No.        &  --    & 5810R \\
			
			Nozzle No.           &  --    & 1393 \\
			
			Propeller diameter D &    m   & 0.10 \\
			
			Number of blades Z   &   --   & 4 \\
			
			Pitch P/D            &   --   & 1.00 \\
			
			100\% RPM            & rev/min& 1059 \\
			\hline
		\end{tabular}
	\end{center}
\end{table}
%
For the validation purpose, a set of five experiments related to straight thrusters mentioned in the reference, \cite{RN959} has been considered in the present study. The detail conditions have been listed in Table~\ref{MARIN_tests}. The advance coefficient in the table is defined as
\begin{equation}
J=\frac{U_\infty}{nD}
\label{ac}
\end{equation}
$U_\infty$ is the inlet current flow and $n$ is the rotation rate of the propeller, corresponding to the RPM (revolutions per minute). 
\begin{table}[htbp]
	\caption{STRAIGHT THRUSTER PHYSICAL EXPERIMENTS in MARIN JIP\cite{RN959}}
	\begin{center}
		\label{MARIN_tests}
		\begin{tabular}{c c c c c}
			\hline
			Test & Advance & Current & $ $ & inflow\\
			No. & coefficient & $ $ & RPM & angle\\
			$ $ & $ J $ & $ U_\infty (m/s) $ & $n(rev/min)$ & $\beta$\\
			\hline
			OW/S/1 & BP & 0 & 1059 & $0^\circ$ \\
			
			OW/S/2 & BP & 0 & 741 & $0^\circ$ \\
			
			OW/S/3 & 0.2 & 0.353 & 1059 & $0^\circ$ \\
			
			OW/S/4 & 0.2 & 0.353 & 1059 & $30^\circ$ \\
			
			OW/S/5 & 0.2 & 0.353 & 1059 & $45^\circ$ \\
			\hline
		\end{tabular}
	\end{center}
\end{table}

\section{Numerical Details and Validation}
\subsection{Setup and meshing}
For the Bollard Pull condition, according to \cite{RN959}, zero inflow velocity is not an ideal condition to initialize the numerical simulation, the initial inlet flow is approximately 0.05 m/s based on the advance coefficient $J_{BP}=0.028$ following \cite{RN959}'s suggestion. Meanwhile, the zero inflow velocity has also been tested in this study and negligible discrepancy is found related to the hydrodynamic force. The thruster surface which includes propeller blades, hub and nozzle were set as no-slip wall boundary. The initial and farfield $\tilde{\nu}$ is set to $3\nu$.  

\begin{figure}[htbp]
	\centering
	\includegraphics[width=3.34in]{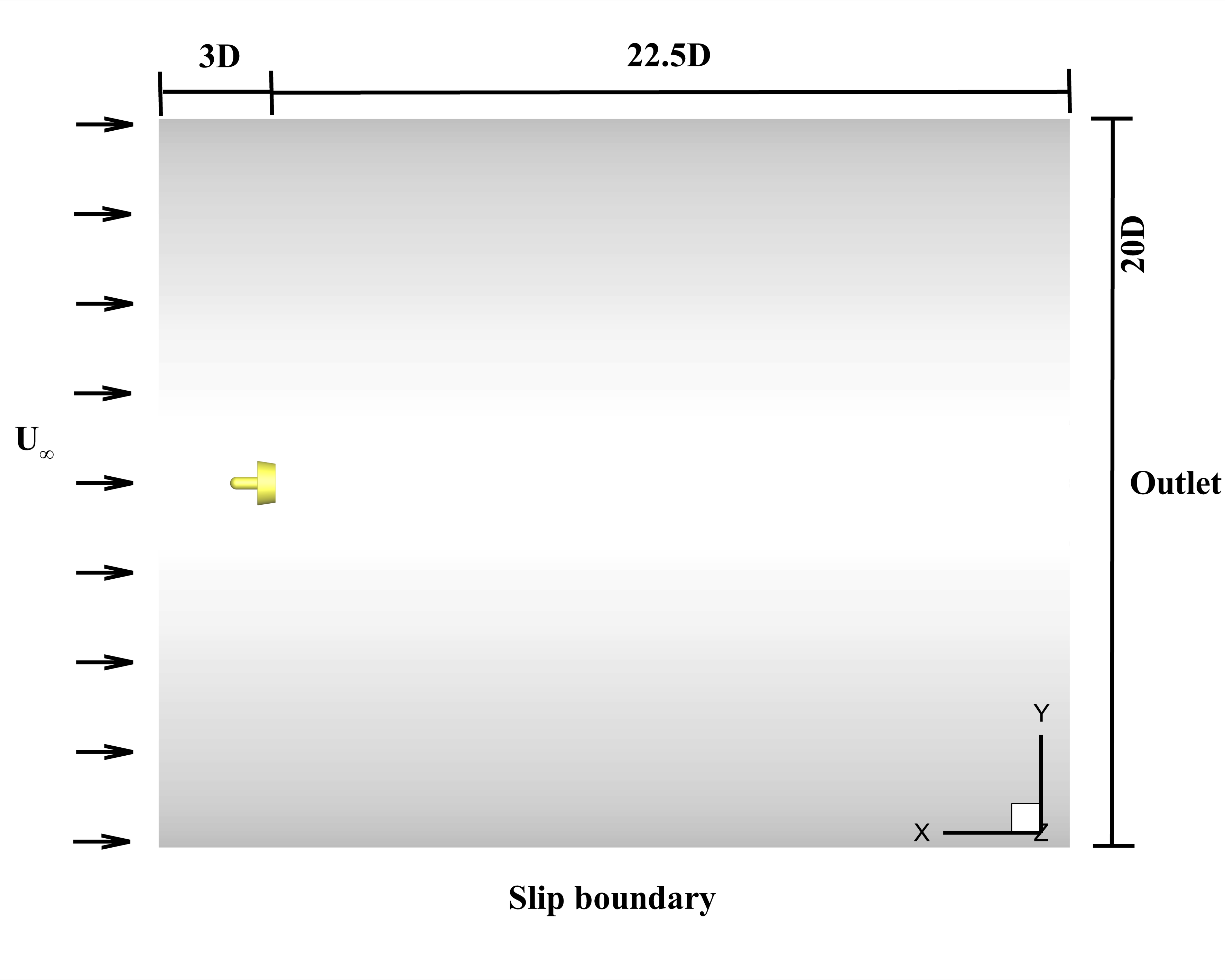}
	\caption{Computational domain of ducted propeller setup.}
	\label{comput_domain}
\end{figure}

\begin{figure}[htbp]
	\centering
	\includegraphics[width=3.34in]{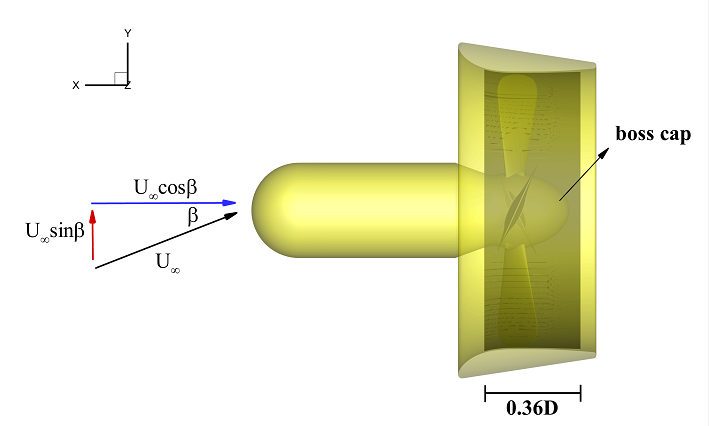}
	\caption{Ducted propeller under oblique flow. The rotor region is indicated by gray color.}
	\label{obliqueflow}
\end{figure}

\begin{figure}[htbp]
	\centering
	\includegraphics[width=3.34in]{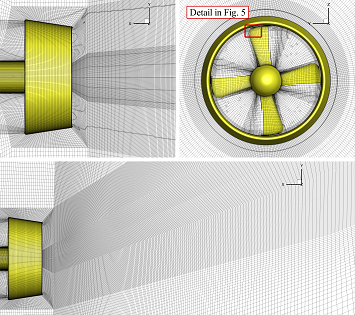}
	\caption{Computational mesh in the vicinity of ducted propeller simulation subjected to axis flow and oblique flow. The close-up mesh in the gap is shown in Fig.~\ref{gap_mesh}.}
	\label{mesh_display}
\end{figure}

The computational domain is a 3D cylindrical shape block as shown in Fig.~\ref{comput_domain}. The length from the edge of the computational domain to the center of ducted propeller in the upstream side is 3D, the downstream side of the ducted propeller extends until 22.5D, the diameter of the cylinder domain is 20D. The inflow velocity and angle is defined by the sketch in Fig.~\ref{obliqueflow}. The computational domain is discretized using block structured grids, including propeller blades and nozzle. The surface structured mesh of ducted thruster considering inflow angle is shown in Fig.~\ref{mesh_display}. It should be noted that in order to simplify the mesh region with the structured mesh, the handle (strut) adhering to the thruster was not meshed in this study.

\begin{figure}[htbp]
	\centering
	\includegraphics[width=3.34in]{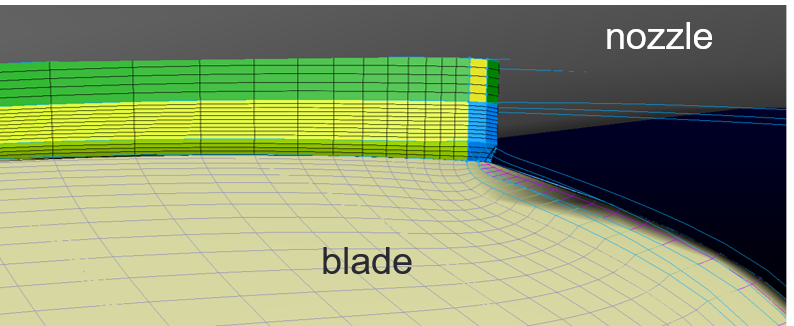}
	\caption{Close-up mesh in the gap between blade tip and nozzle.}
	\label{gap_mesh}
\end{figure}

The same stationary and rotating zones are set for both steady-state and transient simulations.  For the rotating zone, it is set to be a trapezoidal body around the blade, the length in the X-axis of the region is approximately 0.36D (gray area in Fig.~\ref{obliqueflow}), which is just enough to cover the four propeller blades and boss cap, and the outer surface of rotational zone is at the center line of the gap between the nozzle inside surface and the propeller blade tip, as shown in Fig.~\ref{gap_mesh}. The surface of the rotational zone follows the shape of nozzle inside the surface. The rest of the flow field of the computational domain is treated as the ``stator" domain.  

\subsection{Validation}

To evaluate the numerical uncertainty corresponding to different mesh densities, series of MRF steady state simulations were conducted with different mesh sizes, which are listed in Table~\ref{grids_compare}. The physical experiment conducted by MARIN measured the propeller thrust $T_{p}$, the torque $Q_{p}$, and the force on the nozzle ${T_{n}}$. The ratio of propeller thrust to total thrust ${T_{p}}+{T_{n}}$ is provided as a comparable parameter $\tau$ with experimental results, which is listed in Table~\ref{grids_compare}. It can be seen that the ratio does not change dramatically with different mesh densities, which states that the grids generated in this study are adequate to produce reasonable results from a practical viewpoint. The MRF steady state simulation is relatively fast and affordable, therefore, the remaining cases conducted in this study are based on MESH5.

\begin{table}[htbp]
	\caption{MESH SENSITIVITY STUDY}
	\begin{center}
		\label{grids_compare}
		\begin{tabular}{c c c c c}
			\hline
			$ $ & Total & Rotational & $    $ & Difference \\
			$ $ & cell  & region     & $\tau$ & $\%$        \\
			$ $ & $    $  & cells      & $     $& $          $\\
			\hline
			
			OW/S/1   &       &       & 0.5267 &      \\
			
			MESH1 & 6.25M & 0.69M & 0.5089 & 3.38 \\
			
			MESH2 & 13.3M & 1.16M & 0.5110 & 2.98 \\
			
			MESH3 & 17.3M & 2.13M & 0.5126 & 2.68 \\
			
			MESH4 & 23.8M & 2.80M & 0.5161 & 2.01 \\
			
			MESH5 & 41.3M & 7.55M & 0.5186 & 1.54 \\
			
			\hline
		\end{tabular}
	\end{center}
\end{table}

\begin{figure}[htbp]
	\centering
	\includegraphics[width=3.34in]{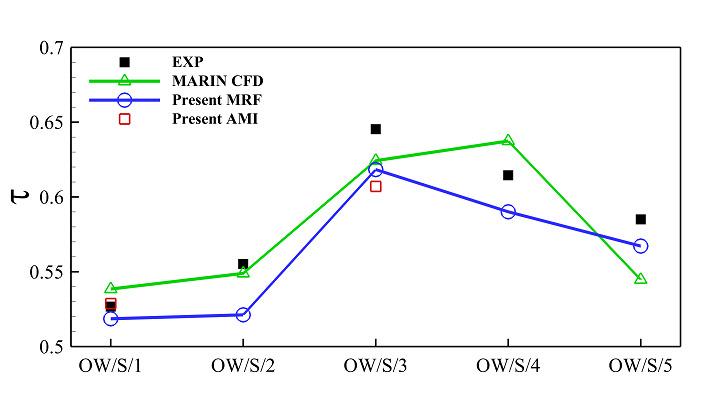}
	\caption{Validation of present MRF and AMI simulation results against experimental and numerical data from the literature.}
	\label{results_compare}
\end{figure}

The numerically simulated results compared with the experimental data as well as the numerical results conducted by MARIN\cite{RN959} are plotted in Fig.~\ref{results_compare}. In general, a good agreement is achieved, which is not surprising since the quasi-steady MRF method and transient sliding mesh method have been proved effectively in performance calculations related to turbomachinery by many other researchers. Only the present MRF and AMI results is slightly lower than the last four cases of experimental results, which maybe due to the current simulation is not involved the effect of the strut on ducted propeller. 

\section{Results and Discussion}
In this study, the numerical results with MRF simulations are mainly focused on the forces and moments generated by the propeller and the nozzle, which are relevant from a practical standpoint. The pressure distribution on the blade and the nozzle with respect to different oblique flow angles are presented in this section. 
The standard non-dimensional representation for propeller thrust $T_p$, torque $Q_p$, nozzle thrust $T_n$ and efficiency $\eta$ are calculated as follows:
\begin{align}
\notag & K_{TP}=\dfrac{T_p}{\rho n^2 D^4}, \quad K_{QP}=\dfrac{Q_p}{\rho n^2 D^5}, \\ & K_{TN}=\dfrac{T_n}{\rho n^2 D^4}, \quad \eta=\dfrac{J ( K_{TP}+K_{TN})}{2 \pi K_{QP}}.
\label{KQ}
\end{align}  
The pressure field developed on the blade and the nozzle is described in terms of pressure coefficient $C_p$, which is calculated as follows:
\begin{equation}
C_p=\frac{p-p_\infty}{\frac{1}{2} \rho U^2_{tip}}
\label{C_p}
\end{equation}
where $p_\infty$ is the static pressure in the free stream and $U_{tip}$ is the velocity of the blade tip ($U_{tip}=n \pi D \simeq 5.54 m/s$). 

\subsection{Propeller with and without nozzle}
\begin{figure}[thpb]
	\centering
	\subfigure[Thrust force]{\label{KT}\includegraphics[width=1.6in]{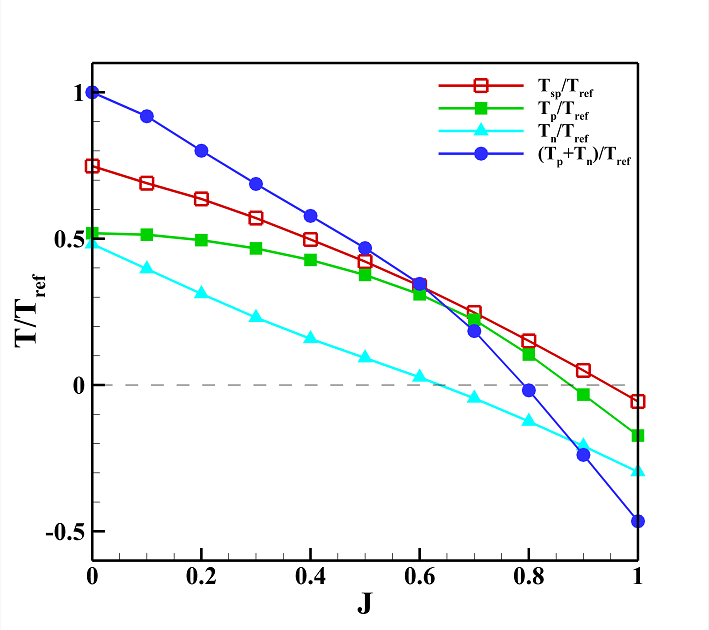}}
	\subfigure[Torque and efficiency]{\label{KQ_eta}\includegraphics[width=1.6in]{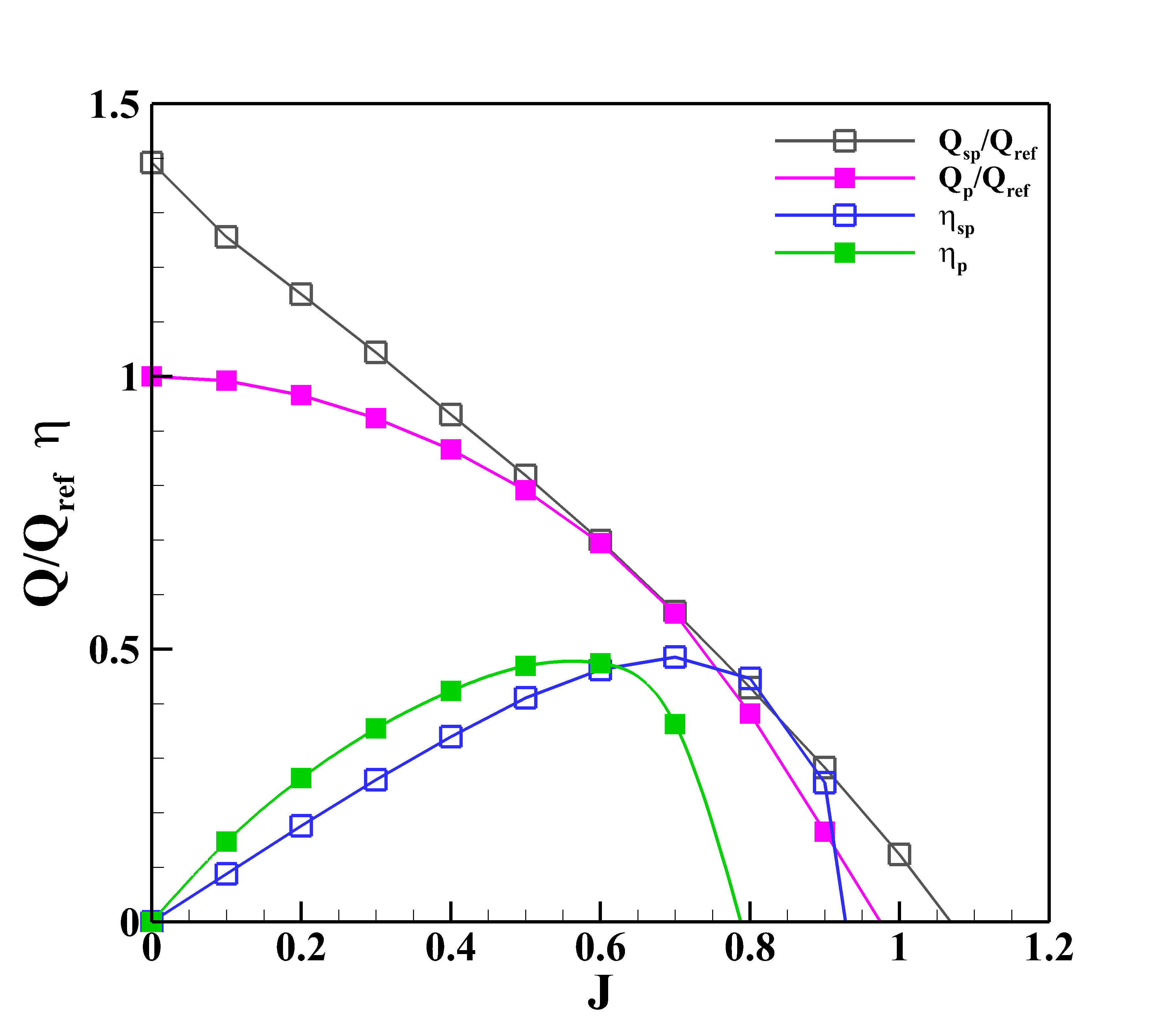}}
	\caption{Comparison of performance characteristics of non-ducted and ducted propellers in open water.} \label{propeller_with_without_nozzle}
\end{figure}

For the sake of completeness, the model thruster has been first simulated with and without nozzle in pure axial flow conditions under a wide range of advance coefficients. The results of the numerical simulation are briefly summarized in this section. The thrust force of propeller without nozzle (non-ducted propeller) is denoted by $T_{sp}$. For the propeller with nozzle (ducted propeller), the thrust generated by propeller and nozzle are represented by $T_p$ and $T_n$ respectively. These quantities are plotted in Fig.~\ref{KT}. For better comparison, all the thrust forces have been divided with the total thrust $(T_p+T_n)$ of the ducted propeller under Bollard Pull (BP) condition $T_{ref}$. 
Firstly, for the ducted propeller, the propeller and nozzle (green line and blue in Fig.~\ref{KT}) have an equal contribution to the total thrust force at BP condition. Although the propeller component of thrust is lower than that of the non-ducted propeller (red line in Fig.~\ref{KT}), the total thrust generated by the ducted propeller is 33\% larger than that of the non-ducted propeller at BP condition. Secondly, with the increase of the advance coefficient, the thrust force decreases for the non-ducted propeller and the ducted propeller, in which, thrust generated by nozzle shows a linear decrease and becomes negative after $J\approx0.6$, others show a parabolic drop and eventually become negative. 
The behavior of the torque and efficiency is plotted in Fig.~\ref{KQ_eta}. The torque shows a similar trend as the thrust force. The required torque for the non-ducted propeller (black line in Fig.~\ref{KQ_eta}) is 50\% more than that of the ducted propeller (purple line) at BP condition. In terms of the efficiency, it is observed that the ducted propeller has a better efficiency when $J<0.6$. The above observation are consistent with\cite{RN1259}. it is evident that the ducted propeller has a better performance than the non-ducted propeller in high loaded condition (small J) due to the nozzle around the propeller. In the next section, the performance of the ducted propeller in oblique flow condition will be discussed systemically.

\subsection{Oblique flow condition}
In order to investigate the performance of the ducted propeller with different incidence angles in a systematical manner, the oblique flow experiments OW/S/4 and OW/S/5 in Table~\ref{MARIN_tests} have been extended to the full range of advance coefficients corresponding to the ducted propeller simulation. Other inflow angles such as $15^\circ$ and $60^\circ$ have also been simulated with advance coefficient below 0.6. The global loads of the ducted propeller in oblique flow will be first discussed in the X,Y and Z axis directions, followed by the pressure distribution on each blade and nozzle and the velocity field near the ducted propeller and the blade tip.

\subsubsection{Global fluid loads}
\begin{figure}[thpb]
	\centering
	\subfigure[Propeller component and nozzle component thrust force.]{\label{3degTpTn}\includegraphics[width=1.6in]{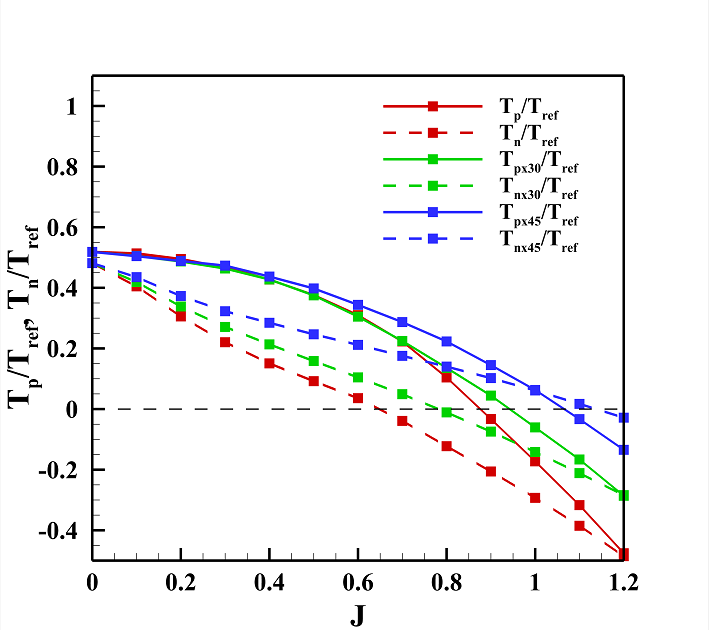}}
	\subfigure[Ducted propeller total thrust and propeller component torque.]{\label{3degTQ}\includegraphics[width=1.6in]{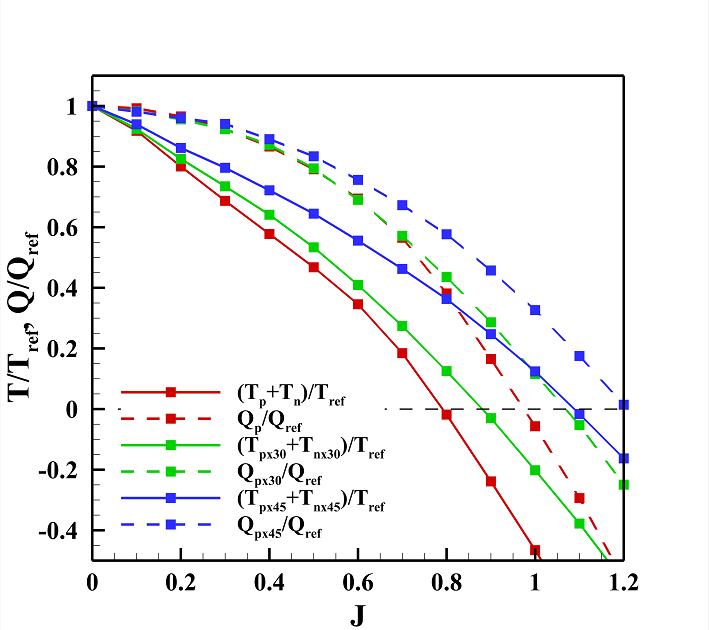}}
	\caption{Ducted propeller open water characteristics under $0^\circ$, $30^\circ$ and $45^\circ$ inflow angles.} \label{3deg}
\end{figure}

\begin{figure}[htbp]
	\centering
	\includegraphics[width=3in]{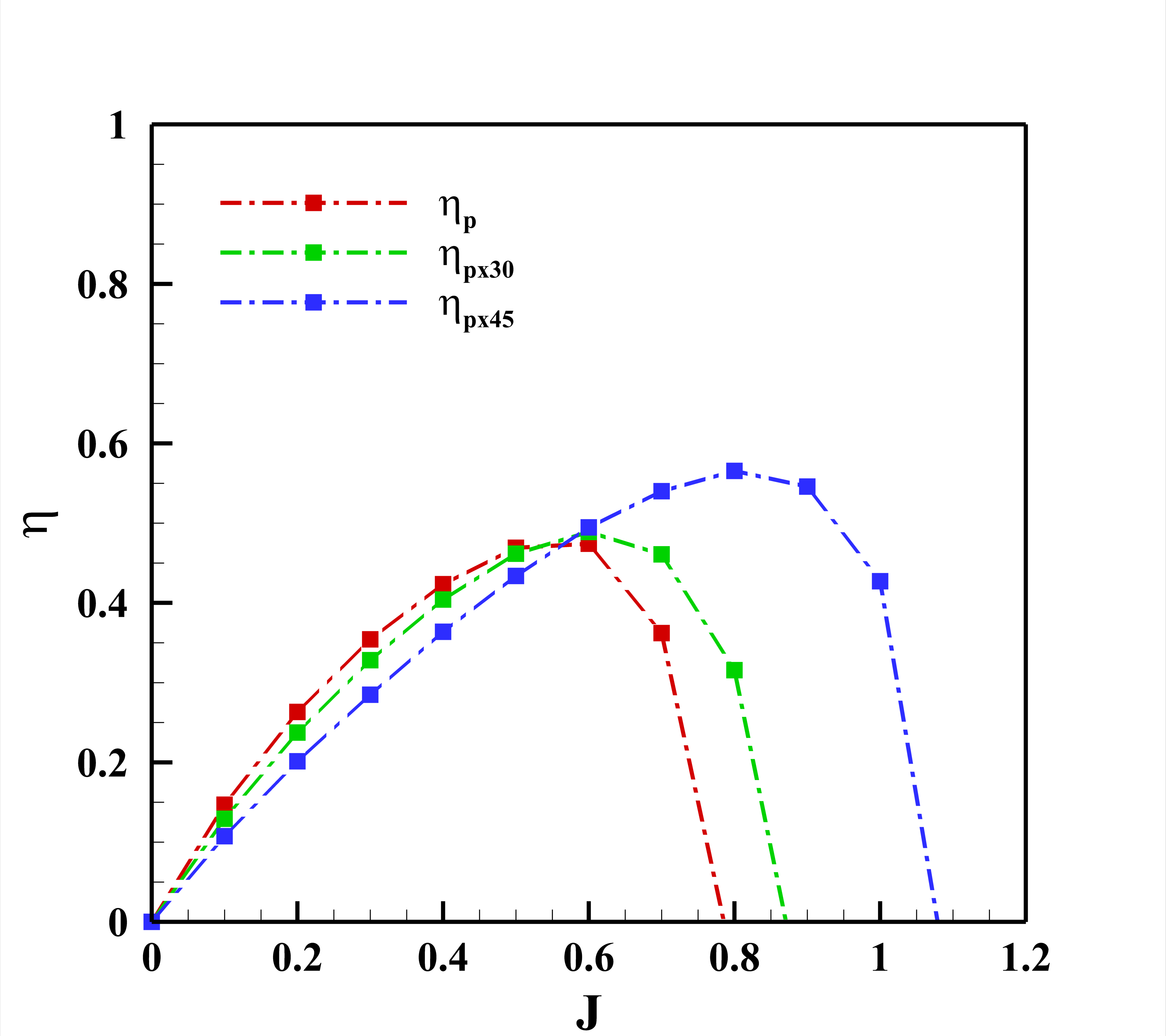}
	\caption{Ducted propeller efficiency under $0^\circ$, $30^\circ$ and $45^\circ$ inflow angle.}
	\label{3deg_eta}
\end{figure}

\begin{figure}[htbp]
	\centering
	\includegraphics[width=3.34in]{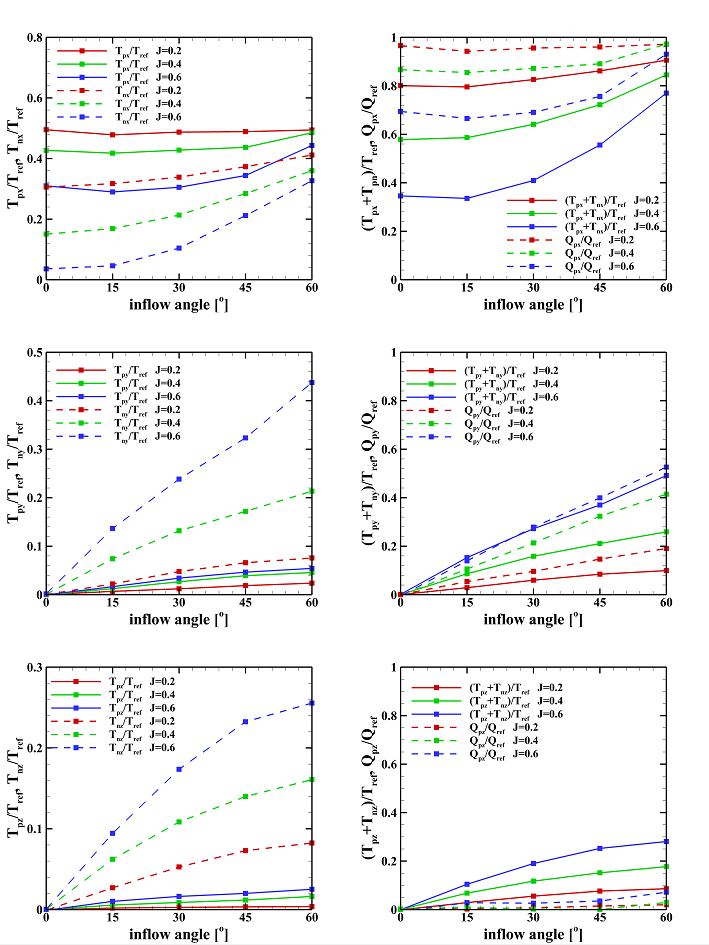}
	\caption{Ducted propeller load for three representative directions.}
	\label{5degTQxyz}
\end{figure}

Along the shaft axis, the open water performance of the ducted propeller under $0^\circ$, $30^\circ$ and $45^\circ$ angles is plotted in Fig.~\ref{3deg}. A similar trend is observed in the variation of the ducted propeller's thrust and torque as compared to that of the ducted propeller in the uniform flow. For the nozzle, it is seen that the generated thrust decreases with increasing advance coefficient. Moreover, the thrust seems to be higher for large oblique angle at a particular advance coefficient (Fig.~\ref{3degTpTn}). For the propeller, an asymptotic behavior is observed for low advance coefficients while at high $J$, the thrust produced by the $45^\circ$ oblique flow case is higher compared to other oblique angles considered in this study. It can be inferred that in highly loaded condition, i.e., when $J<0.4$, the propeller performance is almost independent of the oblique flow angle. In Fig.~\ref{3degTQ}, the total thrust generated by the thrust and nozzle, and the torque produced by the propeller are shown for different oblique angle cases. All the quantities tend to decrease with increase in the advance coefficient.
The variation of the propeller efficiency with the advance coefficient is depicted in Fig.~\ref{3deg_eta}. It is observed that in the highly loaded condition ($J<0.4$), the efficiency slightly decreases with increase in the oblique angle. At high $J$, the propeller with $45^\circ$ oblique angle has a higher efficiency compared to other cases.
To gain further insight into the performance curves of the ducted propeller, several tests were carried out at different oblique angles ($0^\circ$-$60^\circ$) with three values of the advance coefficient (0.2, 0.4 and 0.6). The thrust and the torque generated were further recorded in the three different directions $X$, $Y$ and $Z$ respectively. These results are summarized in Fig.~\ref{5degTQxyz}. Some of the peculiar findings from the plot are: first, for a particular advance coefficient, the thrust and the torque increases with the increasing oblique angle. Second, the rate of increase in the thrust and the torque with the oblique flow angle is higher for larger advance coefficient. Third, at a fixed oblique inflow angle, the thrust and the torque are larger for the lower advance coefficient in the X direction, while this trend reverses in the Y and Z directions. Apart from the general observations, the torque in the starboard (Y) direction which acts as a pitching moment consists of a significant percentage of the propeller torque. However, negligible yawing (Z) moment is observed for the different oblique flow angles. 

\subsubsection{Pressure on blades and nozzle}
\begin{figure}[htbp]
	\centering
	\includegraphics[width=3.34in]{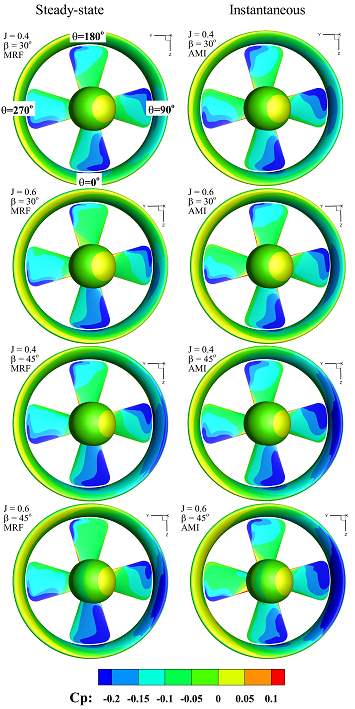}
	\caption{Pressure distribution on suction side for steady-state and transient computations.}
	\label{Cpss}
\end{figure}

\begin{figure}[htbp]
	\centering
	\includegraphics[width=3.34in]{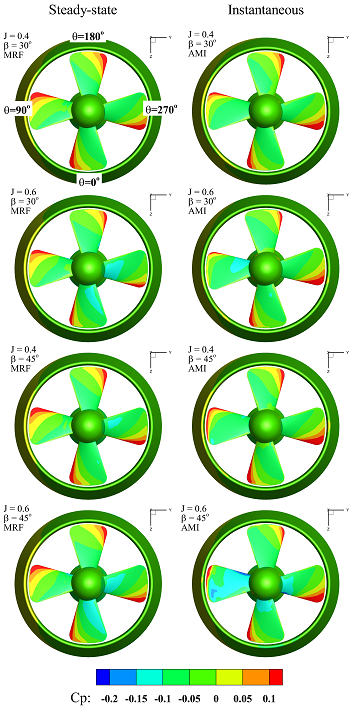}
	\caption{Pressure distribution on pressure side for steady-state and transient computations.}
	\label{Cpps}
\end{figure}

The pressure distribution on the suction side and the pressure side of the ducted propeller with two different values of the advance coefficient and the oblique flow angle are shown in Fig.~\ref{Cpss} and Fig.~\ref{Cpps} respectively. The pressure distribution is visualized by evaluating the pressure coefficient given by Eq.~(\ref{C_p}). The MRF and AMI results are also compared in Fig.~\ref{Cpss} and Fig.~\ref{Cpps}. It is observed that the gradients of the pressure distribution cover a larger area for MRF simulation compared to AMI results suggesting that the MRF simulations may be over-estimating the effects of the oblique angle flow on the pressure. For the suction side of the propeller, a large low pressure area is observed at the blade at $\theta=0^\circ$ compared to other blades. An asymmetric pressure distribution is also noticed for $\theta=90^\circ$ and $\theta=270^\circ$. Moreover, for a constant advance coefficient, an increase in the oblique angle leads to an increase in the pressure distribution area at $\theta=0^\circ$ and $\theta=90^\circ$ and the inner side of the duct (or the nozzle). While on the pressure side of the propeller, an asymmetric pressure distribution is observed almost for all the blades (see Fig.~\ref{Cpps}).

\begin{figure}[htbp]
	\centering
	\includegraphics[width=3.34in]{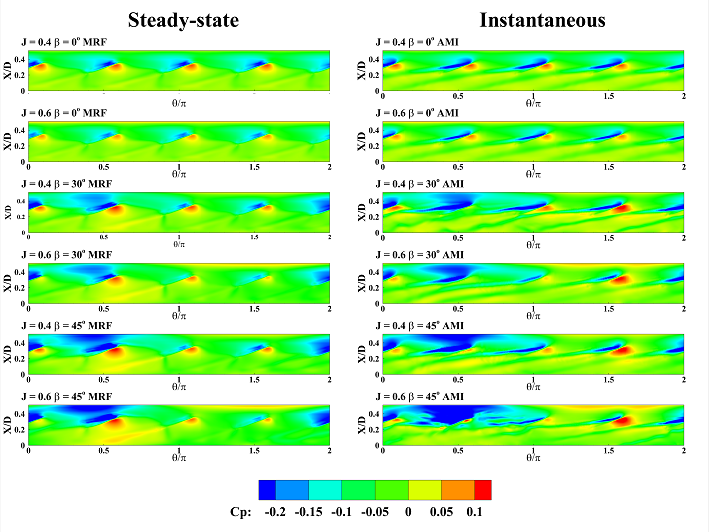}
	\caption{Pressure distribution on the inside surface of nozzle for steady-state and transient computations.}
	\label{Cpni}
\end{figure}

\begin{figure}[htbp]
	\centering
	\includegraphics[width=3.34in]{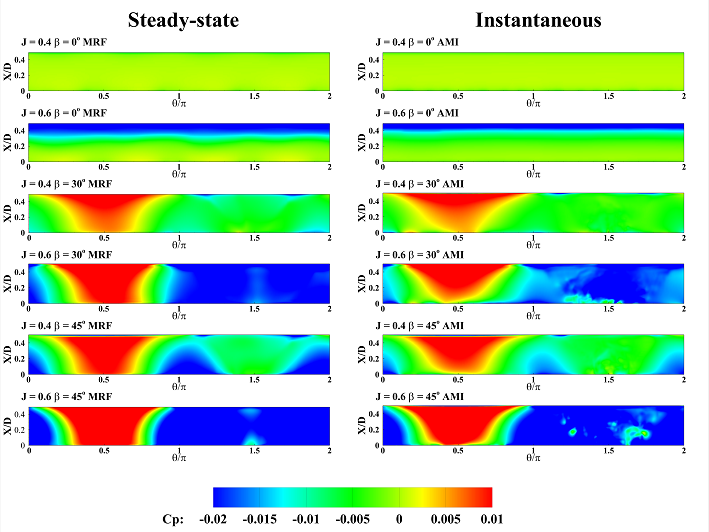}
	\caption{Pressure distribution on the outside surface of nozzle for steady-state and transient computations.}
	\label{Cpno}
\end{figure}

%
The pressure distribution on the inner and the outer surface of the duct (or the nozzle) is depicted in Fig.~\ref{Cpni} and Fig.~\ref{Cpno} respectively. This is carried out by unfolding the cylindrical duct in two-dimensions. Due to the oblique nature of the inflow, a region of low and high pressure at the suction and the pressure side of the nozzle respectively is observed at $\theta=90^\circ$. The outer surface of the nozzle (Fig.~\ref{Cpno}) is generally unaffected by the oblique flow angle. A high pressure area is seen at $\theta=90^\circ$. 

\subsubsection{Pressure blade sections}
\begin{figure}[htbp]
	\centering
	\includegraphics[width=3.34in]{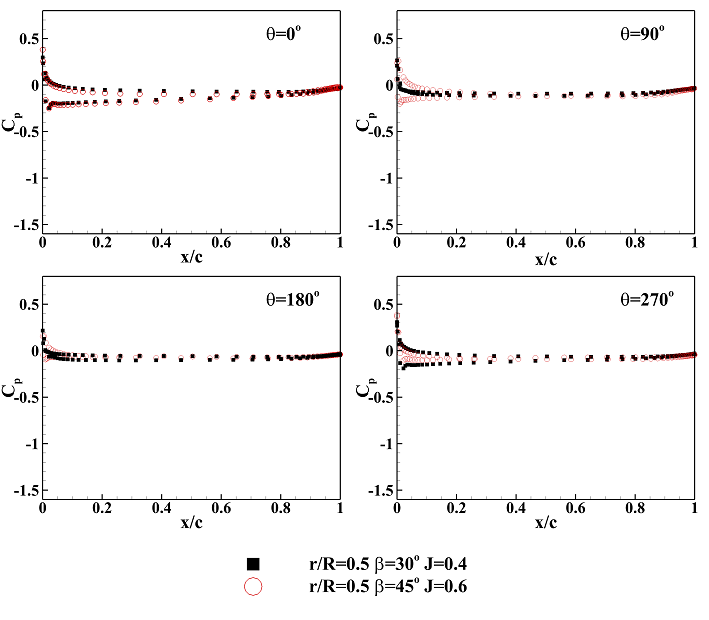}
	\caption{Transient solution of $C_p$ at blade section $r/R=0.5$ for four positions.}
	\label{c05R}
\end{figure}

\begin{figure}[htbp]
	\centering
	\includegraphics[width=3.34in]{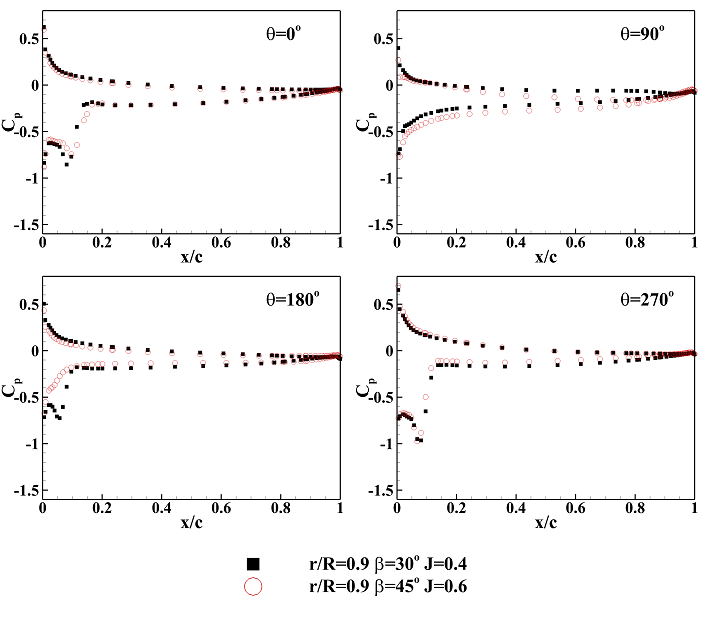}
	\caption{Transient solution of $C_p$ at blade section $r/R=0.9$ over four representative positions.}
	\label{c09R}
\end{figure}

\begin{figure}[htbp]
	\centering
	\includegraphics[width=3.34in]{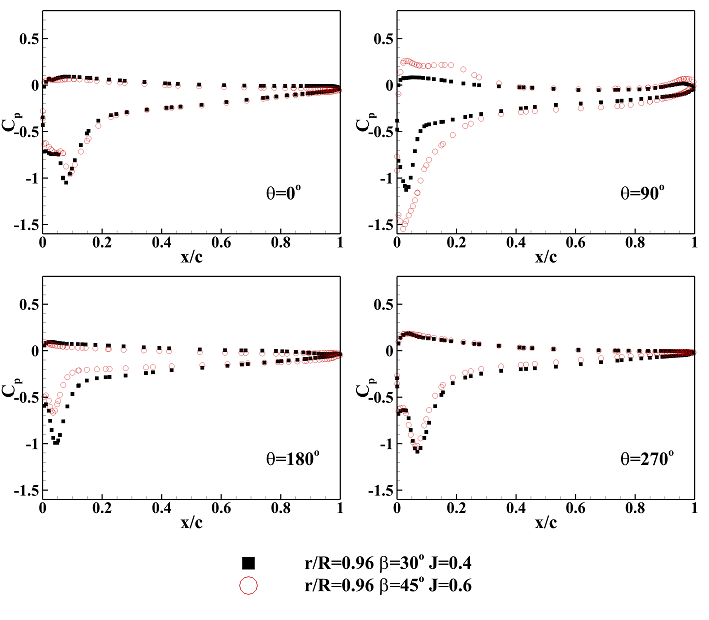}
	\caption{Transient solution of $C_p$ at blade section $r/R=0.96$ over  four representative positions.}
	\label{c096R}
\end{figure}

To provide an insight about the pressure imbalance on the blade sections under the oblique angle flow, the pressure coefficient is plotted along the blade section at $r/R \in [0.5, 0.9, 0.96]$ corresponding to the root, maximum chord and the tip section of the blade in Fig.~\ref{c05R}, Fig.~\ref{c09R} and Fig.~\ref{c096R} respectively. It is observed that for all the three $r/R$ cases considered, the pressure drop is the steepest at the blade directly exposed to the oblique flow ($\theta=90^\circ$) at its leading edge (small $x/c$) for smaller oblique flow angle. While on all the other cases, the trend of the pressure coefficient is quite similar. 

\subsubsection{Flow fields around blade}
\begin{figure}[htbp]
	\centering
	\includegraphics[width=3.34in]{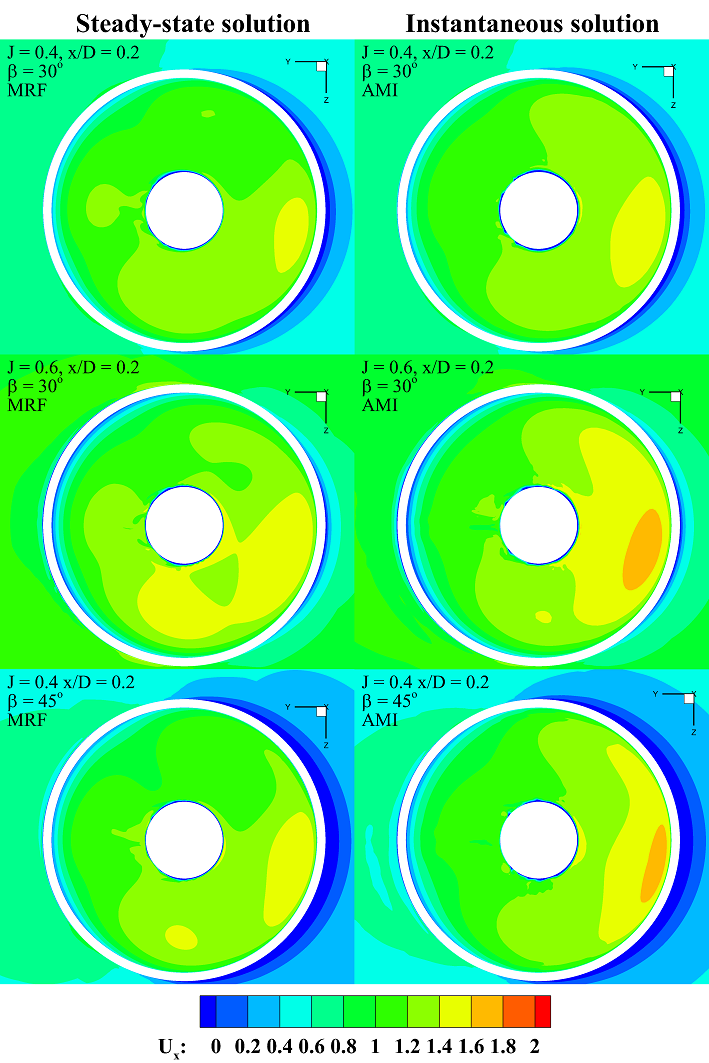}
	\caption{Ux velocity parallel to the propeller plane (X/D=0.2). Left:steady-state solution, right: instantaneous solution.}
	\label{Ux02D}
\end{figure}

\begin{figure}[htbp]
	\centering
	\includegraphics[width=3.34in]{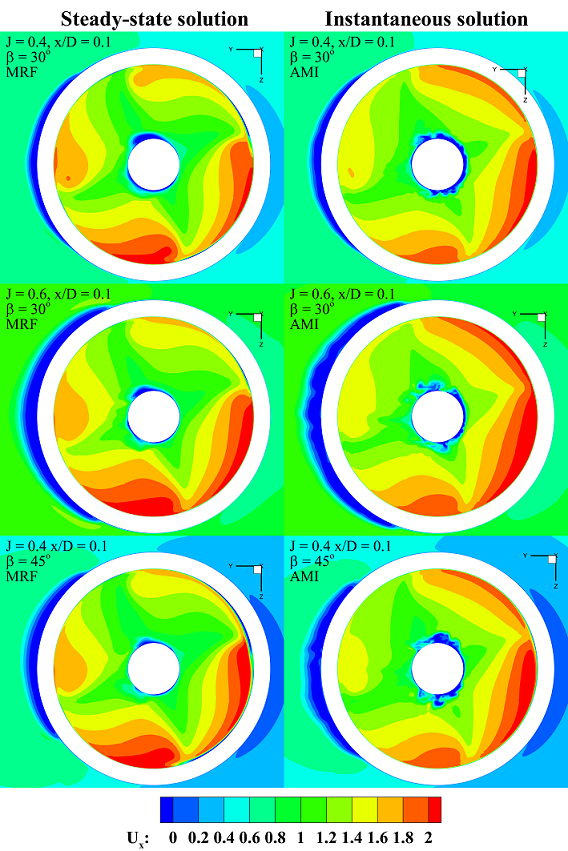}
	\caption{Streamwise $U_x$ velocity parallel to the propeller plane (X/D=0.1). Left: steady-state, right: instantaneous.}
	\label{Ux01D}
\end{figure}

\begin{figure}[htbp]
	\centering
	\includegraphics[width=3.34in]{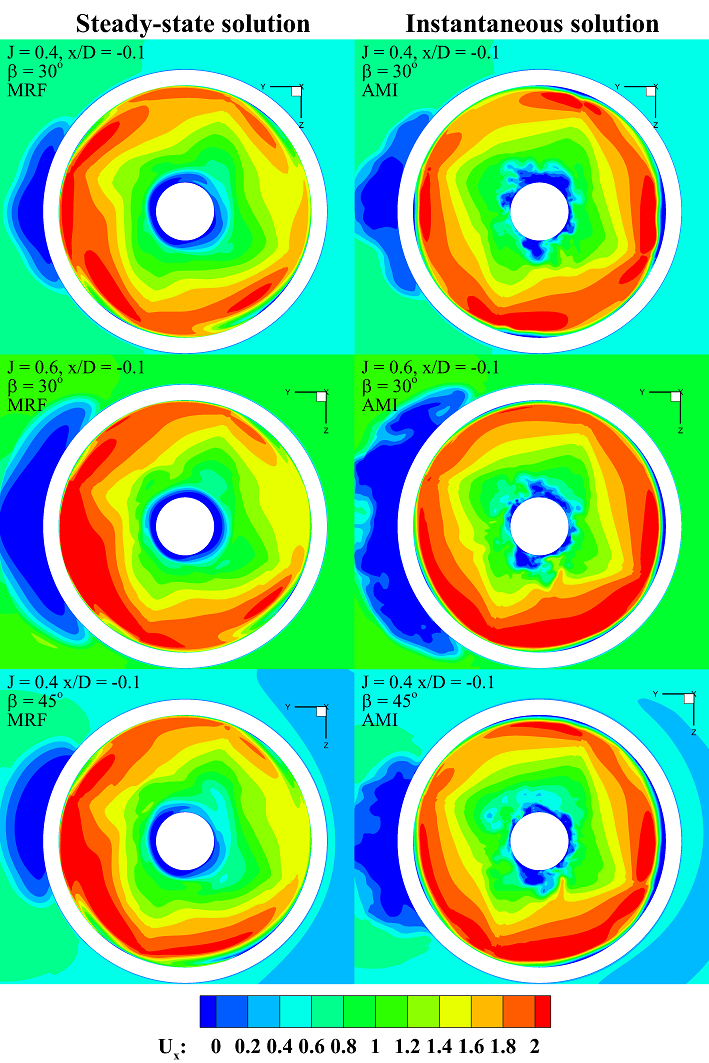}
	\caption{Streamwise $U_x$ velocity parallel to the propeller plane (X/D=-0.1). Left: steady-state, right: instantaneous.}
	\label{Ux-01D}
\end{figure}

\begin{figure}[htbp]
	\centering
	\includegraphics[width=3.34in]{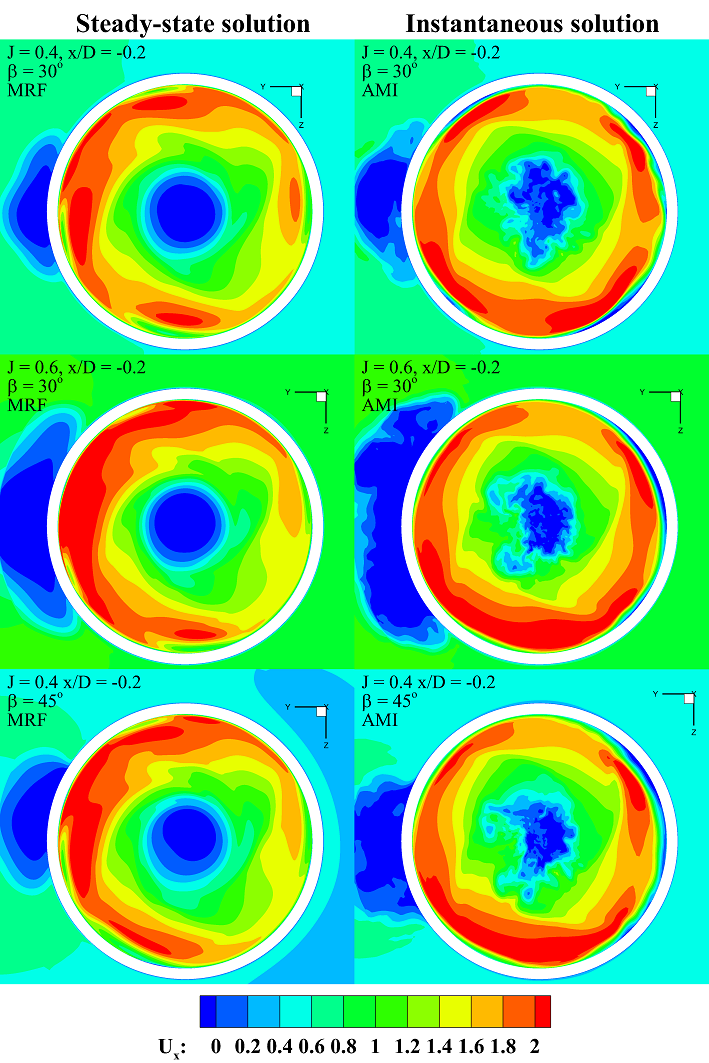}
	\caption{Streamwise $U_x$ velocity parallel to the propeller plane (X/D=-0.2).Left: steady-state solution, right: instantaneous solution.}
	\label{Ux-02D}
\end{figure}

Flow field around the propeller is crucial to understand the force generation process by the propeller. The cross-sectional contours of the axial velocity for the upstream side of the propeller at $x/D \in [0.2, 0.1]$ are shown in Fig.~\ref{Ux02D} and Fig.~\ref{Ux01D} respectively. The flow contours appear similar for both the MRF and AMI simulations. However, more transient effects and high velocity regions are captured by the AMI simulation. The increase of the oblique inflow angle tends to increase the low velocity area outside the nozzle (Fig.~\ref{Ux02D}). For the downstream side of the propeller, the contour plots for the axial velocity at $x/D \in [-0.1,-0.2]$ are plotted in Fig.~\ref{Ux-01D} and Fig.~\ref{Ux-02D} respectively. The unsteady velocity components are captured here by the AMI simulation. In this case, with the increase in the advance coefficient, the low velocity area around the nozzle is enlarged.

\begin{figure}[htbp]
	\centering
	\includegraphics[width=3.34in]{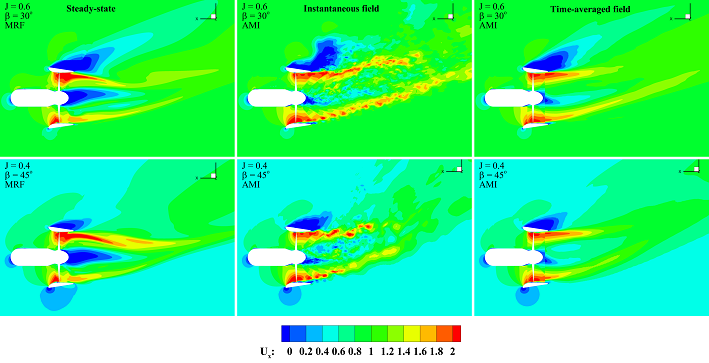}
	\caption{Streamwise $U_x$ velocity in the horizontal plane. Left: steady-state solution, right: instantaneous solution.}
	\label{fdUx}
\end{figure}

The wake flow of ducted propeller is also shown in Fig.~\ref{fdUx}. The figure includes the results from the MRF (steady state), AMI (transient) and the time averaged AMI simulations. The time averaging is carried out over a time period of 10 revolutions of the propeller. It is evident that the transient AMI simulations are able to capture the transient wake flow patterns, providing a detailed information about the flow physics of the turbulent wake. Therefore, transient AMI simulations with sliding mesh is the better choice to investigate the unsteady flow field and the turbulent wake characteristics behind the ducted propeller. 

\section{CONCLUSION}
In this paper, the RANS-based steady state simulation coupled with the MRF method and the LES-based transient simulation coupled with the sliding structured mesh have been applied for open and ducted propeller simulations in open water under axis flow and different oblique flow conditions. Both methodologies were validated using the available model test data and a reasonable agreement is obtained in terms of force predictions. The results show that for the axis flow, the ducted propeller requires a lesser torque for more thruster generation compared with the open propeller counterpart, but the performance reduced much faster than the open propeller with the increase of advance coefficient. For the oblique flow condition, with the increase of advance coefficient and the inflow angle, the nozzle force increases faster than the propeller force and the oblique flow has more effect on the suction side than the pressure side of the propeller blade. From the analysis of the pressure distribution along the propeller blade, despite the nozzle of the ducted propeller absorbs much effect from oblique flow, the imbalance between the different blade and position still exists. The steady state and the transient simulations are also compared with respect to the pressure distribution on the blades and nozzle as well as the flow field. Some inconsistency is observed according to pressure distribution, both the steady state and transient results show limited effect of oblique flow on the blade due to the presence of nozzle. However,  the transient simulation with the sliding mesh is necessary for studying unsteady flow features and dynamics accurately.  
%
\begin{acknowledgment}
This work done at the Keppel-NUS Corporate Laboratory was supported by the National Research Foundation, National University of Singapore (NUS) and Keppel Offshore and Marine Technology Centre (KOMtech).
The authors specifically would like to acknowledge the help and support by ir. Hans Cozijn and Dr. ir. Arjen Koop from MARIN for providing us with the geometry and useful communication.
We also acknowledge the support from National Supercomputing Centre, Singapore (NSCC) for  the computational resource. Discussion with Mr. Vaibhav Joshi and Mr. Guan Mengzhao are greatly appreciated.
\end{acknowledgment}

%
% ...

\bibliographystyle{asmems4}

\bibliography{asme2e}

%

\end{document}